# Large-area topological wireless power transfer

Luyao Wan[1], Han Zhang[1], Yunhui Li[2], Yaping Yang[1], Hong Chen[1], and Zhiwei Guo[1,*]

[1]School of Physics Science and Engineering, Tongji University, Shanghai 200092, China

[2]Department of Electrical Engineering, Tongji University, Shanghai 201804, China

*Correspondence should be addressed to: 2014guozhiwei@tongji.edu.cn

**Abstract:** Topological wireless power transfer (WPT) technologies have attracted considerable interest due to their high transmission efficiency and robustness in coupled array configurations. However, conventional periodic and quasi-periodic topological chains exhibit limited adaptability in complex application scenarios, such as large-area simultaneous multi-load charging. In this work, we experimentally demonstrate a large-area topological defect state by constructing a gapless chain of uniformly coupled resonators at the interface of two topologically distinct Su-Schrieffer-Heeger (SSH) configurations. This topological defect state exhibits strong localization at multiple target sites, enabling efficient and concurrent wireless power delivery to spatially distributed loads. Furthermore, the unique wavefunction distribution enhances robustness against positional variations, ensuring stable energy transfer despite fluctuations in device placement. The proposed large-area topological framework offers fundamental insights into harnessing diverse topological states for advanced WPT applications, particularly in scenarios demanding spatial flexibility and multi-target energy delivery.

**Keywords**: Topological phase transition; defect state; Near-field coupling; Wireless power transfer.

## Introduction

Wireless power transfer (WPT) has garnered significant research attention in recent years due to its wide-ranging applications in consumer electronics, robotics, and electric vehicles [1–4]. Conventional magnetic inductive WPT systems, however, suffer from limited transmission efficiency due to the rapid exponential decay of power with increasing distance and coil misalignment [5,6]. Moreover, severe magnetic field leakage in inductive WPT schemes can generate electromagnetic interference, potentially disrupting nearby electronic devices or biological systems. Another major challenge lies in maintaining optimal power transfer efficiency under variable load conditions, particularly in high-power applications [7–9]. The breakthrough introduction of magnetic resonance coupling in 2007 significantly extended the transmission range of near-field WPT by enabling efficient energy transfer between resonantly tuned transmitter and receiver coils [10]. However, this approach requires precise frequency tracking, necessitating dynamic adjustments to compensate for variations in transmission distance [11]. To mitigate these frequency instability issues, researchers have explored advanced optimization strategies, such as nonlinear saturation gain control [13–15] and high-order parity-time- (PT-) symmetry [16–19]. While these methods improve transmission efficiency and system robustness, they remain ineffective in complex energy transfer scenarios, such as robotic arms and electric insulator string configurations, where longer transmission distances and greater structural deformations are required [20]. To further extend the transmission range, researchers have introduced relay coils to form a Domino chain, effectively overcoming the size constraints of individual coils [21–23]. However, this approach is highly sensitive to positional disturbances, as variations in coupling strength between coils lead to significant efficiency degradation. Furthermore, the cumulative parasitic resistance introduced by additional coils becomes increasingly non-negligible, further compromising overall transmission efficiency. These inherent limitations fundamentally restrict the practical deployment of Domino chain-based systems in complex environments.

The integration of topological photonics with near-field WPT has recently enabled robust energy transmission through unique state distributions [24–29]. For instance, researchers have implemented Su-Schrieffer-Heeger (SSH) chains with coil resonators, which supports topologically protected edge states [30, 31]. These states exhibit strong immune against positional perturbations, ensuring stable energy delivery. However, because energy is predominantly localized at the terminal sites of the chain, this approach restricts power transfer to a single receiver [25–27], making it unsuitable for simultaneous multi-load WPT. Recent advances in topological photonics have sparked growing interest in large-area topological edge states associated with topological phase transition, which offer exceptional robustness against structural disorder while enabling spatially extended field confinement [32–35]. Unlike conventional localized edge modes, these states facilitate electromagnetic wave manipulation over extended geometries while preserving topological protection—an essential feature for applications in low-loss multi-channel energy transfer [36]. Experimental realizations in non-Hermitian photonic crystals [37–39], acoustics [40–42], and waveguide platforms [43, 44] have further demonstrated the feasibility of large-area topological transport. Given these advantages, incorporating the large area topological physics mechanism into WPT presents a promising avenue for enhancing efficiency and scalability.

In this work, we present a pioneering study on topological large-area defect states (LADS) for multi-load WPT. By inserting a gapless chain of uniformly coupled resonators at the interface of two topologically distinct SSH configurations, we experimentally demonstrate the emergence of LADS with a uniform state distribution across multiple target sites. This unique energy localization enables simultaneous multi-load charging, which we conduct a visual experimental demonstration using Light Emitting Diode (LED) lamps as representative loads. Furthermore, we systematically investigate the robustness of LADS against external perturbations, highlighting its resilience to positional misalignment and structural disorder. The findings of this work provide valuable insights into the potential of topological physics in advancing WPT technologies, particularly for applications requiring spatially distributed energy delivery with enhanced stability.

## Results

**LADS in a composite topological chain**. Figures 1(a)-1(c) show the topological transition of SSH chain with three different configurations. $\kappa_1$ and $\kappa_2$ represent the coupling between adjacent atoms, where $\kappa_1$ denotes the strong coupling strength and $\kappa_2$ corresponds to the weak coupling strength. When the intracell coupling exceeds the intercell coupling, the open-boundary energy spectrum of the chain exhibits two distinct energy bands separated by a finite band gap, as shown in Fig. 1(a). As the intracell coupling strength equals to the intercell coupling strength, a complete closure of the band gap is observed in the energy spectrum. As the intracell coupling becomes weaker than the intercell coupling, the band gap reopens and is accompanied by a characteristic band inversion, signifying a topological phase transition in the system. The topological invariant (i.e., Zak phase) of topological trivial and nontrivial phases are marked near the bands. In Fig. 1(d), a simple topological defect is introduced at the interface between two chains with distinct configurations, exhibiting weak coupling strengths to the edge atoms of both chains [45]. This heterojunction dimer chain realizes a defect state at zero energy. After expanding the interfacial atom into a uniformly coupled chain, the originally localized edge state is expanded to a LADS, with energy uniformly distributed across the odd sites.

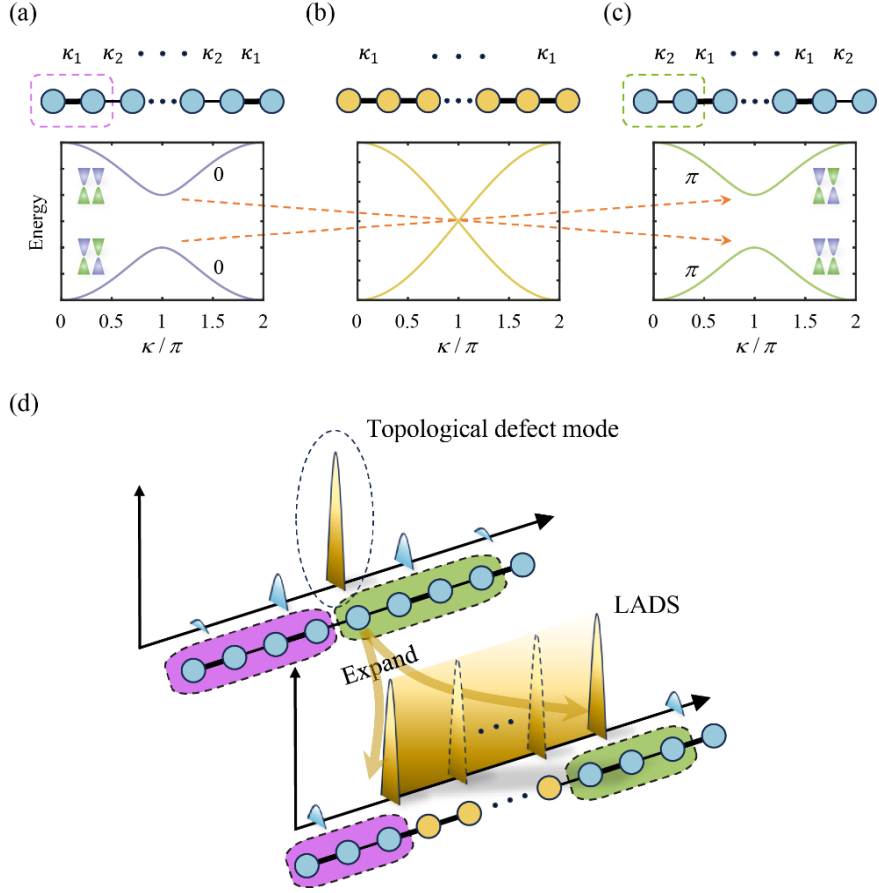

**Fig. 1. Schematic of the topological LADS and enhanced homogeneous field.** (a)-(c) Topological phase transition of the dimer chain. The dashed box represents the unit cell containing two sites, where the intra-cell and inter-cell coupling strength are $\kappa_1$ and $\kappa_2$, respectively. The thick (thin) line between two sublattices represents strong (weak) coupling strength with $\kappa_1 > \kappa_2$. A topological transition occurs in (b) with the uniform coupling strength $\kappa_1$ between adjacent site. The band gap reopens as the coupling strength changes. (d) Schematics of LADS expanded from the topological defect state in the heterojunction dimer chain. The topological defect state appears at the interface between two topological distinguished chains. After expanding the single interface site to a chain with uniform coupling strength, the edge state is expanded in a large area.

Figure 2(a) depicts a chain consists of 15 coil resonators. The parameters of each coil are maintained identical. The resonant frequency is 358 kHz, with a capacitance of 22 nF, and the diameter is standardized at 10 cm. The coupling strength between two resonators depends on their separation distance. Through experimental measurements, we extracted the relationship between the coupling strength and separation distance of coils, as shown in Fig. 2(b). The coupling strength between two coils corresponds to half the frequency splitting in the spectrum. The measured results and fitted curve are represented by red dots and solid line, respectively. The fitted function between coupling strength $\kappa$ and distance $d$ is $\kappa = 79 e^{-d/2.29} + 6.08$ [46]. Guided by this quantitative correlation, precise control of coupling between adjacent coils can be effectively implemented. Then we employed a small loop antenna connected to a vector network analyzer (VNA) to detect the density of states (DOS) of the system based on the principle of near-field probing [47]. Owing to the substantial frequency detuning between the probe and the coil resonators' frequency, the loop antenna could be approximated as a high-impedance probing element. The antenna is positioned at the center of the test coil, so energy can be injected into the system through the antenna. By measuring the reflection coefficient $S_{11}$, the local density of states (LDOS) can be obtained using $1 - {S_{11}}^2$. The DOS of the topological chain is derived by summing and averaging the LDOS contributions of all coils.

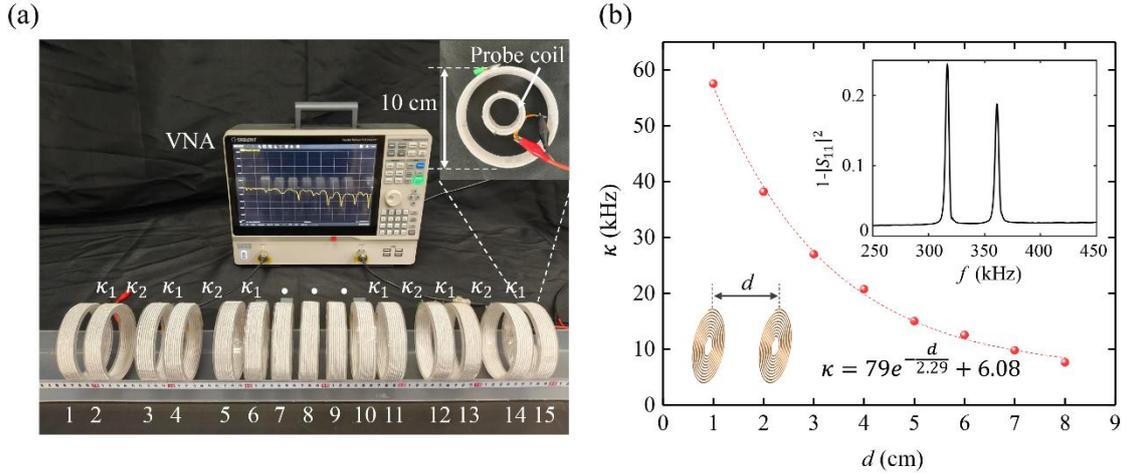

**Fig. 2. Experimental setup and the coupling strength $\kappa$ as a function of the distance *d* between two adjacent coils.** (a) The topological chain composed of 15 coil resonators. The diameter and the resonant frequency of the coil are 10 cm and 358 kHz, respectively. The loaded capacitor is 22 nF. The inset shows the probe coil used for measuring the reflection spectrum. (b) The coupling strength at different distances and fitted curve. The red dots are the measured results and the dashed line is the fitted curve. The fitting function is $\kappa = 79e^{-d/2.29} + 6.08$. The inset shows the measured frequency spectrum when *d* = 4 cm.

We initially construct a heterojunction dimer chain comprising nine coil resonators to study the topological defect state [45]. Figure 3(a) shows the schematic of the composite dimer chain composed of two topological distinguished chains. By calculating the eigenvalues of the system, we can find an isolated mode at the zero energy, which corresponds to the topological defect state, as shown in Fig. 3(b). The topological defect mode is marked by the arrow. From the calculated LDOS distribution of the topological defect state in Fig. 3(c), it can be found that the energy is mainly concentrated on the resonators at the interface. Moreover, the measured DOS spectrum and LDOS distributions are shown in Figs. 3(d) and 3(e), respectively. An isolated state at the bandgap can be easily determined from the DOS spectrum, and the measured LDOS results are meet well with the calculated ones.

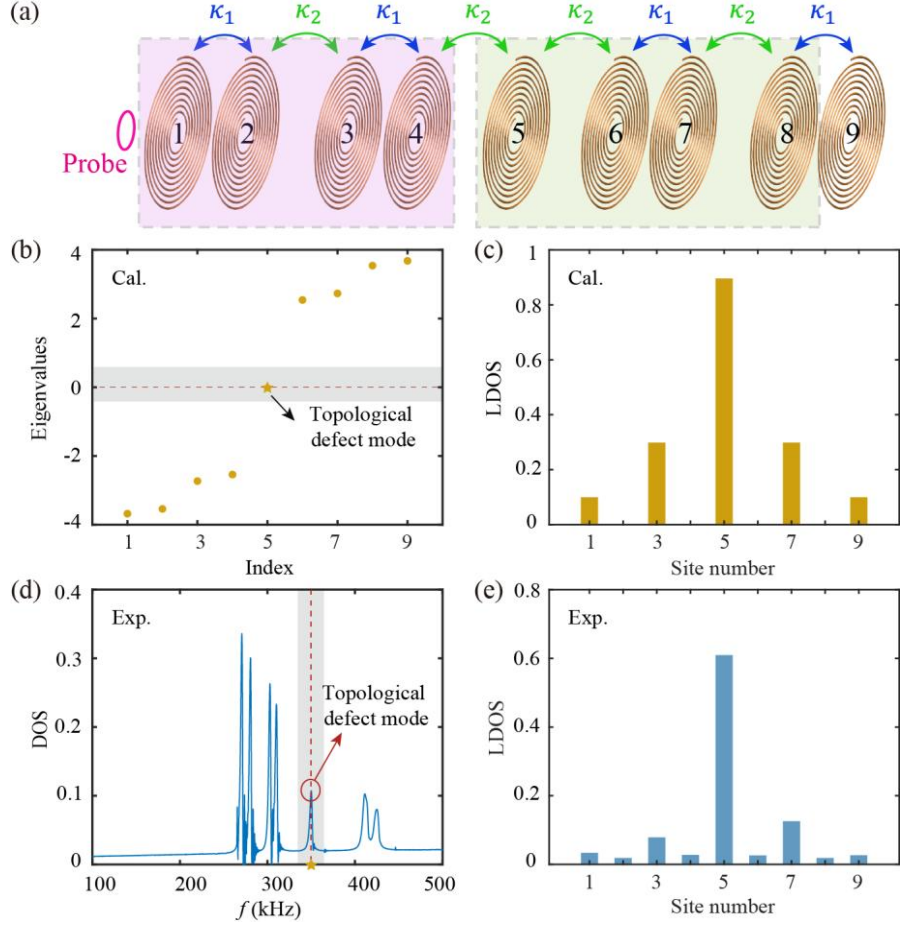

**Fig. 3. Topological defect state in heterojunction dimer chain with $\kappa_1/\kappa_2=3$.** (a) Schematic of the composite dimer chain composed of two topological distinguished chains. (b) The corresponding eigenvalues spectra of the heterojunction topological chain. The topological defect state is marked by the arrow. (c) Calculated LDOS distribution of topological defect state. (d) Measured DOS spectrum of the heterojunction topological chain. The topological defect state at 350 kHz is marked by the circle. (e) Measured LDOS distribution of topological defect state.

We further expand the single coil at the interface site to seven uniformly coupled coils, where the coupling strength between each coil is $\kappa_1$. The expanded dimer chain is schematically shown in Fig. 4(a). From the calculated eigenvalues in Fig. 4(b), we can see the isolated LADS in the bandgap. Especially, the LADS is highly localized at multiple sites instead of one fixed site, as shown in Fig. 4(c). Similar to Fig. 3, we also measured DOS spectrum and LDOS distributions of the LADS, as shown in Figs. 4(d) and 4(e), respectively. The LADS (i.e., 360 kHz) exists in the bandgap and is circled by the red circle. At this time, the energy is evenly distributed on the coils at odd sites. The slight differences between experimental results and theoretical calculations stem from sample construction and testing errors.

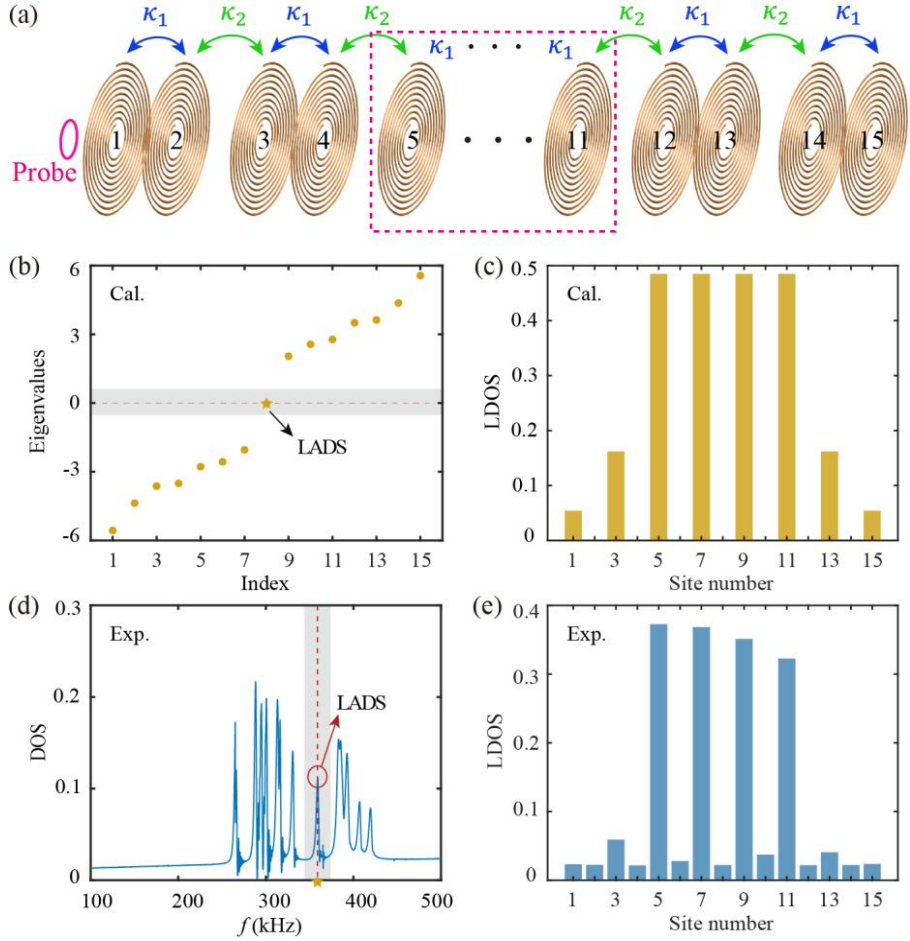

**Fig. 4. Topological LADS in heterojunction dimer chain with $\kappa_1/\kappa_2$ =3.** (a) Schematic of the expanded dimer chain. (b) The real part of the energy spectrum. The zero mode corresponds to LADS. (c) Calculated LDOS distribution of LADS. (d) Measured DOS spectrum of the expanded dimer chain. The frequency of the LADS is 358 kHz because of the next-nearest-coupling and errors generated during coil sample preparation. One bulk state (260 kHz) is also marked by the circle. (e) Measured LDOS distribution of LADS.

Subsequently, we employ LED indicators to visualize the state distribution of LADS. The chain is excited by a high-power signal source (AG Series Amplifier, T&C Power Conversion), with a source coil positioned at the edge of the uniformly coupled coil resonators to establish LADS. Each resonant coil is added with an LED indicator through non-resonant coil. When the magnetic field strength on a resonant coil exceeds a critical threshold, the corresponding LED indicator is illuminated. At the working frequency of LADS, the LED indicators at the 5, 7, 9 and 11 sites are lit up, while all other indicators remain dark, as illustrated in Fig. 5. As a result, the design of LADS for multi-load charging is intuitively demonstrated.

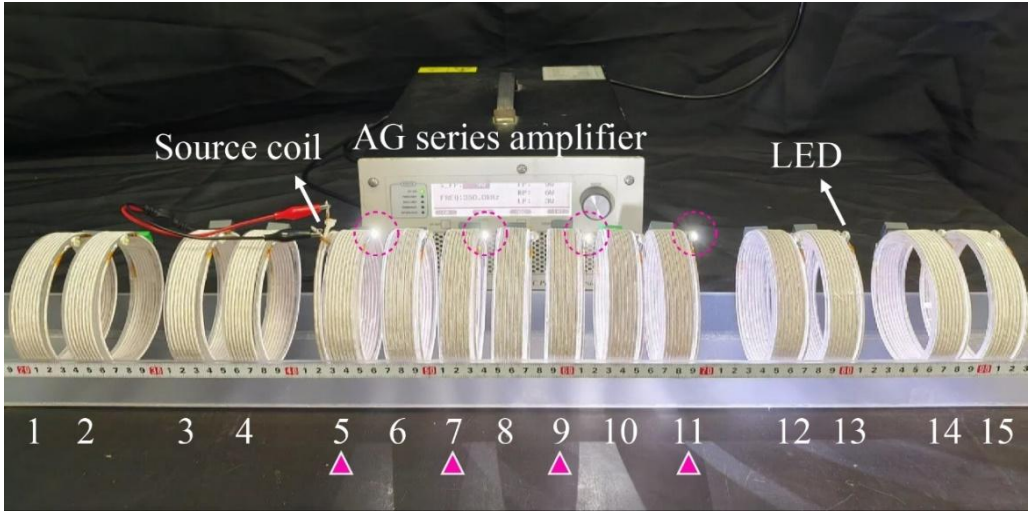

**Fig. 5.** Experimental demonstration of the 3 W WPT for lighting LED lamps. The source coil is placed at the edge of the uniformly coupled coils. The LED lamps are mounted on all the resonant coils. At the working frequency, the energy of the LADS is predominantly localized at sites 5, 7, 9, and 11, with corresponding LEDs at these sites being illuminated. In contrast, LEDs at other sites remain dark due to the relatively weak magnetic field.

**Characterization of LADS**. The energy localization degree of LADS varies under different coupling conditions. We calculate LDOS of the system for $\kappa_1/\kappa_2$ ranging from 1 to 6, as shown in Fig. 6(a). The inverse participation ratio (IPR) is defined as $\sum_n|\varphi_n|^4/(\sum_n|\varphi_n|^2)^2$, which serves as a standard metric for quantifying localization [48, 49]. The IPR exhibits a pronounced increasing trend with the progressive enhancement of $\kappa_1$. In the experimental configuration, with $\kappa_2$ fixed at 15 kHz and $\kappa_1$ incrementally set to 15, 30, 45 and 60 kHz, the energy distribution demonstrates progressive concentration on the uniformly coupled coils as $\kappa_1$ increases, as shown in Fig. 5(b)-5(e), respectively.

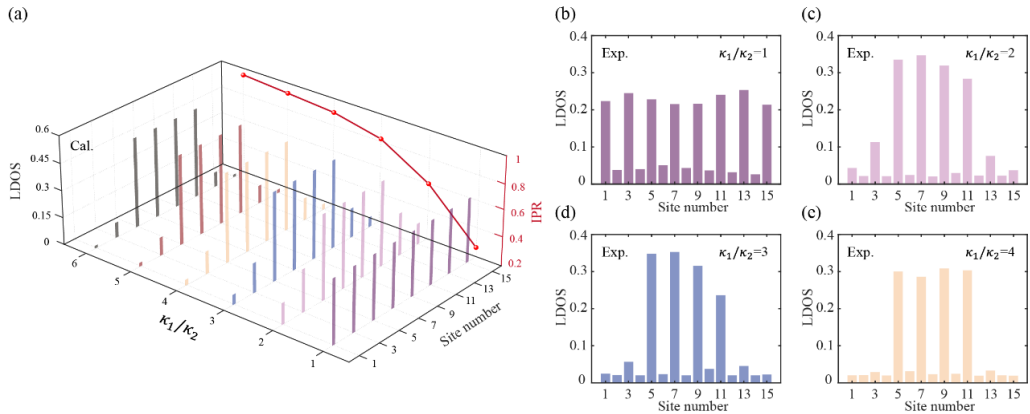

**Fig. 6. LDOS and inverse participation ratio (IPR) with different coupling strength.** (a) Calculated LDOS distributions and IPR. When $\kappa_1/\kappa_2$ =1, the field strength on each unit is equal. As the increasement of $\kappa_1$, the energy is more localized on the expanded units. IPR rises from 0.25 to 0.95. (b-e) Measured LDOS distributions. Here we fix $\kappa_2$ at 15 kHz and the values of $\kappa_1$ are 15 kHz, 30 kHz, 45kHz and 60 kHz, respectively.

**Robustness against perturbations.** We experimentally investigated the robustness of LADS against positional perturbations. A topological chain comprising 13 coil resonators was constructed. At this time $\kappa_1$ is 15 kHz and $\kappa_2$ is 45 kHz, corresponds to 5.5 cm and 1.5 cm, respectively. The perturbations are introduced by randomly varying the distance between adjacent coils. We first generate random numbers within the range of 0 to 1, then multiply by a displacement factor $\Delta d=0.5$ cm to determine the change of the distance between adjacent coils. We select a bulk state for comparison. It can be observed that there was no significant change in the frequencies of both LADS and bulk state after the introduction of perturbations. The LDOS of bulk state and LADS are also measured. There is a significant change in the state distribution of the bulk state, whereas the energy of LADS remains concentrated on the central three coil resonators, only a slight variation in amplitude compared to the unperturbed case. These findings demonstrate that LADS exhibits strong robustness against positional perturbations.

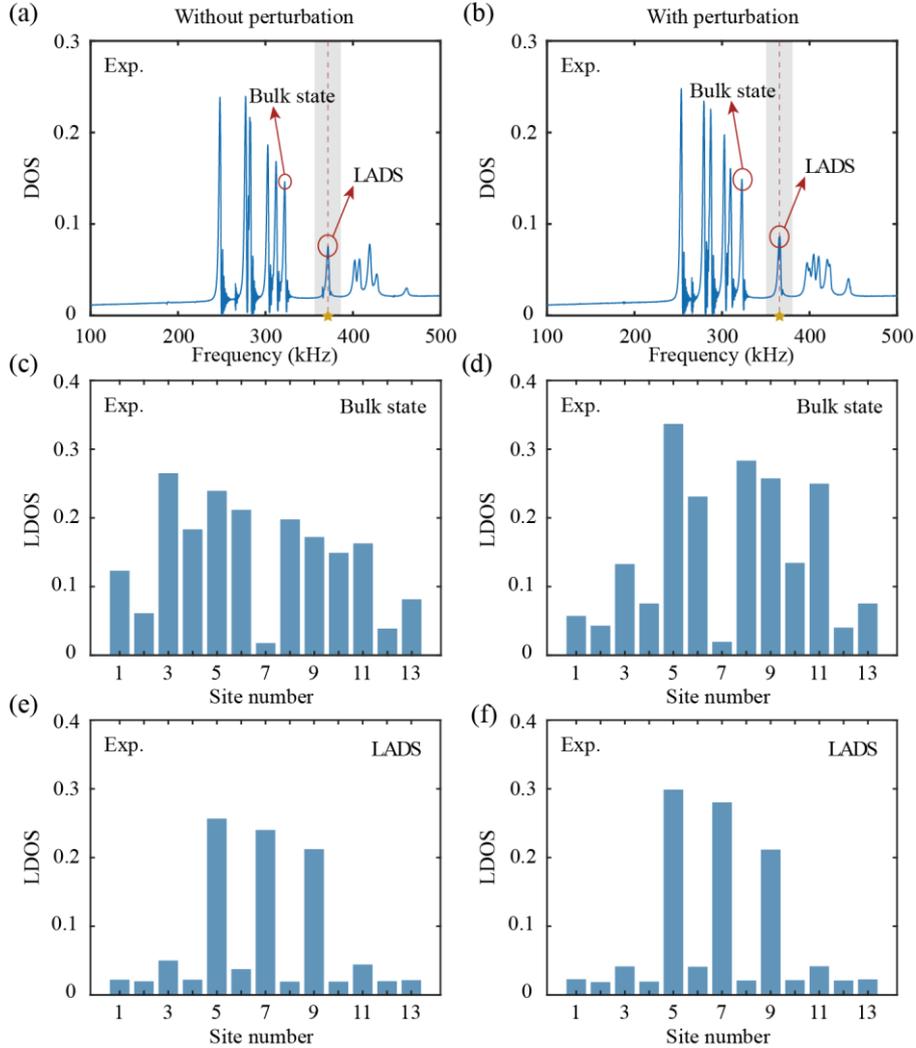

**Fig. 7. Robust LADS in an expanded composite topological chain composed of 13 coils.** (a-b) Measured DOS spectrum without (with) perturbation. (c-d) Measured LDOS distributions of the bulk state without (with) perturbation. (e-f) Measured LDOS distributions of the LADS without (with) perturbation.

## Conclusion

In summary, we have experimentally demonstrated a novel approach to achieving LADS in a one-dimensional topological WPT system based on subwavelength coil resonators. By leveraging the topological properties of SSH chains, we established a robust defect state that enables efficient and simultaneous power delivery to multiple spatially distributed loads. Unlike conventional topological WPT schemes, where energy localization is confined to the system's edges, our approach extends the energy distribution over a broader region while maintaining resilience against positional perturbations and structural disorder. This work provides a foundational framework for multi-load WPT, paving the way for practical implementations in scenarios demanding spatial flexibility and enhanced stability. The demonstrated topological mechanism is not only applicable to planar coil arrays but also holds the potential for higher-dimensional extensions, enabling new possibilities for robust and scalable wireless energy networks.

## Data availability

All the data that support the findings of this study are available from the corresponding authors upon reasonable request.

## Code availability

All the codes that support the findings of this study are available from the corresponding authors upon reasonable request.

## Acknowledgments

This work is supported by the National Key R&D Program of China (Nos. 2021YFA1400602, and 2023YFA1407600), the National Natural Science Foundation of China (Nos. 52477014 and 12374294), the Interdisciplinary key project of Tongji University (No. 2023-1-ZD-02), the Shanghai Science and Technology Commission Project (No. 2021SHZDZX0100), and the Chenguang Program of Shanghai (No. 21CGA22).

## Author contributions

Z. G. and H. C. conceived the idea and supervised the project. L. W. and H. Z. carried out the analytical calculations with the help of Y. L, and Y. Y. L. W. prepared the sample and conducted experimental measurements. L. W. H. Z. and Z. G. wrote the manuscript. All authors contributed to discussions of the results and the manuscript.

## Competing interests

The authors declare no competing financial and non-financial interests.